%% file: paper.tex
\documentclass[conference]{IEEEtran}
\IEEEoverridecommandlockouts

\usepackage{cite}
\usepackage{amsmath,amssymb,amsfonts}
\usepackage{algorithmic}
\usepackage{graphicx}
\usepackage{textcomp}
\usepackage{xcolor}
\usepackage[left=0.673in,right=0.673in,top=0.7in,bottom=0.967in]{geometry}
\def\BibTeX{{\rm B\kern-.05em{\sc i\kern-.025em b}\kern-.08em
T\kern-.1667em\lower.7ex\hbox{E}\kern-.125emX}}

\usepackage[acronym]{glossaries}
\makeglossaries
\input{acronyms}

\begin{document}

\title{Cooperative Deep Reinforcement Learning for Fair RIS Allocation
\thanks{This research was funded in whole or in part by the
Austrian Science Fund (FWF) 10.55776/PAT4490824. For
open access purposes, the author has applied a CC BY public
copyright license to any author accepted manuscript version
arising from this submission.}
}

\author{\IEEEauthorblockN{1\textsuperscript{st} Martin Mark Zan}
\IEEEauthorblockA{\textit{Institute of Telecommunications} \\
\textit{TU Wien}\\
Vienna, Austria \\
martin.zan@tuwien.ac.at}
\and
\IEEEauthorblockN{2\textsuperscript{nd} Stefan Schwarz}
\IEEEauthorblockA{\textit{Institute of Telecommunications} \\
\textit{TU Wien}\\
Vienna, Austria \\
stefan.schwarz@tuwien.ac.at}
}

\maketitle

\begin{minipage}{500pt}
\vspace{-400pt}
\centering{
\footnotesize{© 2026 IEEE. Personal use of this material is permitted. This is the author's version of the work accepted for publication in International Symposium on Modeling and Optimization in Mobile, Ad hoc, and Wireless Networks (WiOpt 2026).}}
\end{minipage}

\begin{abstract}
The deployment of \glspl{ris} introduces new challenges for resource allocation in multi-cell wireless networks, particularly when user loads are uneven across base stations. In this work, we consider RISs as shared infrastructure that must be dynamically assigned among competing base stations, and we address this problem using a simultaneous ascending auction mechanism.

To mitigate performance imbalances between cells, we propose a fairness-aware collaborative multi-agent reinforcement learning approach in which base stations adapt their bidding strategies based on both expected utility gains and relative service quality. A centrally computed performance-dependent fairness indicator is incorporated into the agents’ observations, enabling implicit coordination without direct inter-base-station communication. 

Simulation results show that the proposed framework effectively redistributes RIS resources toward weaker-performing cells, substantially improving the rates of the worst-served users while preserving overall throughput. The results demonstrate that fairness-oriented RIS allocation can be achieved through cooperative learning, providing a flexible tool for balancing efficiency and equity in future wireless networks.
\end{abstract}


\section{Introduction}
 
\IEEEPARstart{I}{n} the evolution toward 6G wireless networks, intelligent resource management has become a central challenge in interference-limited environments. While advances in spectral efficiency and massive antenna systems have significantly improved peak data rates, ensuring fair and reliable service across users and cells remains a key objective, particularly at the cell edge where propagation conditions are poor and competition for shared resources is most pronounced. 

While cell-edge performance limitations are addressed by techniques such as \gls{comp} \cite{Irram2020}, \cite{Schwarz2014} and cell-free massive MIMO \cite{Nayebi2015,Mendoza2025}, the practical deployment of these approaches is constrained by limited coordination cluster sizes \cite{Ammar2022}, \cite{Mendoza2023}. As a result, full network-wide coordination remains infeasible, and performance degradation persists at the edges of coordination clusters rather than at conventional cell boundaries. The framework considered in this work is therefore complementary to CoMP and cell-free MIMO.

\Acrfullpl{ris} have recently emerged as a promising technology to address these challenges by enabling programmable control of the wireless propagation environment \cite{Zeng2021}. By adjusting the phase responses of nearly passive reflecting elements, RISs provide a cost- and energy-efficient means to enhance desired signal paths and mitigate interference, thereby complementing conventional base station and user equipment capabilities \cite{Msleh2023, Le2021}.  

Despite their potential, practical RIS deployment raises system-level questions regarding placement and efficient allocation among multiple transmitters and users. These challenges are particularly pronounced in multi-cell scenarios, where RISs deployed near cell boundaries can benefit multiple base stations, leading to competition for shared infrastructure.

To address this competition, RISs are modeled as shared resources managed by an independent infrastructure provider and dynamically leased to \glspl{bs} via a market-inspired allocation mechanism. In particular, auction-based allocation provides a scalable and low-complexity alternative to combinatorial optimization approaches \cite{Nisan2001}, while explicitly capturing the strategic interactions among competing base stations. Similar auction formats have been successfully applied in spectrum allocation, where simultaneous ascending auctions enable efficient distribution of scarce resources \cite{Cramton2017, Milgrom2004}. An auction-based RIS allocation mechanism was recently studied in \cite{Schwarz2024} for a multi-operator scenario, demonstrating the viability of this approach in RIS-assisted networks.

Building on this framework, \gls{rl} enables optimized bidding strategies in dynamic and partially observable environments. By learning from repeated auction interactions, RL agents adapt their behavior to target high-value RISs while avoiding inefficient bidding, and have been shown to outperform heuristic approaches in performance–cost trade-offs \cite{Zan2026}. Deep RL has been successfully used to coordinate a central resource provider and multiple competing operators in multi-RIS networks in \cite{Zhang2025}. Auction-based energy markets have been formulated as stochastic games, where reinforcement learning is used to learn effective bidding strategies under uncertainty and competition in \cite{Nanduri2007}.

In contrast to existing work, this paper studies fairness-aware RIS allocation in asymmetric multi-cell scenarios with uneven user distributions. We introduce a performance-dependent fairness indicator into the RL agents’ observations, enabling implicit coordination that favors weaker-performing cells when beneficial. A tunable parameter controls the trade-off between total throughput and equitable resource distribution.

Simulation results show that the proposed cooperative multi-agent RL framework yields a more balanced RIS allocation, significantly improving minimum user rates in overloaded cells and reducing the Atkinson inequality index, while maintaining competitive sum-rate performance. We further demonstrate how fairness settings shape agent behavior and system-level outcomes.

\textit{Notation:} The multi-variate complex Gaussian distribution with mean $\boldsymbol{\mu}$ and covariance matrix $\boldsymbol{C}$ is denoted by $\mathcal{CN}(\boldsymbol{\mu},\boldsymbol{C})$, while the uniform distribution over the interval $[a,b]$ is written as $\mathcal{U}(a,b)$. For a vector $\boldsymbol{x}$, the transpose and Hermitian transpose are $\boldsymbol{x}^T$ and $\boldsymbol{x}^H$, respectively, and the $i$-th element is denoted by $\boldsymbol{x}[i]$. The Euclidean norm of $\boldsymbol{x}$ is $\|\boldsymbol{x}\|$. The cardinality of a set $\mathcal{X}$ is $|\mathcal{X}|$, and the empty set is denoted by $\emptyset$. The expectation operator is written as $\mathbb{E}[\cdot]$, and the phase of a complex number $z$ is given by $\arg(z)$.

\section{System Model}

We consider a multi-cell downlink scenario with $N_{\mathrm{BS}}$ base stations, serving a total of $N_{\mathrm{UE}}$ single-antenna users with the assistance of $N_{\mathrm{RIS}}$ reconfigurable intelligent surfaces. Each BS is equipped with $M_{\mathrm{BS}}$ antennas, while each RIS consists of $M_{\mathrm{RIS}}$ reflecting elements. We consider single-user MIMO, i.e., users are served on orthogonal resources.

\subsection{Channel Model}

We consider both direct and RIS-assisted channels between each BS and each user.  We assume that the direct BS-\gls{ue} link is dominated by \gls{nlos} propagation and is strongly shadowed, which motivates the use of RIS-assisted transmission. The direct channel between user $u$ and BS $b$ is denoted by $\boldsymbol{h}^{\mathrm{direct}}_{u,b} \in \mathbb{C}^{M_{\mathrm{BS}} \times 1}$ and modeled as
\begin{equation}
\boldsymbol{h}^{\mathrm{direct}}_{u,b} = \gamma_{u,b} \boldsymbol{g}_{u,b},
\end{equation}
where $\gamma_{u,b}$ is the path gain, and $\boldsymbol{g}_{u,b} \sim \mathcal{CN}(\boldsymbol{0}, \boldsymbol{I})$ is the fading component. The normalization is chosen such that $\mathbb{E}[\|\boldsymbol{g}_{u,b}\|^2] = M_{\mathrm{BS}}$, yielding $\mathbb{E}[\|\boldsymbol{h}^{\mathrm{direct}}_{u,b}\|^2] = \gamma_{u,b}^2 M_{\mathrm{BS}}$. We consider Rayleigh fading for this link to model strong multipath propagation under NLOS conditions.

For the channel between BS $b$ and RIS $r$, a strong line-of-sight (LOS) component is assumed, as RISs are typically deployed in locations with good visibility to nearby BSs. This directional LOS channel is represented by the matrix $\boldsymbol{H}_{r,b} \in \mathbb{C}^{M_{\mathrm{RIS}} \times M_{\mathrm{BS}}}$ and modeled as
\begin{equation}
\boldsymbol{H}_{r,b} = \gamma_{r,b} \boldsymbol{a}(\psi_{r,b}) \boldsymbol{a}(\theta_{r,b})^T,
\end{equation}
where $\gamma_{r,b}$ is the corresponding path gain, $\boldsymbol{a}(\psi_{r,b})$ denotes the RIS array response vector, and $\boldsymbol{a}(\theta_{r,b})$ is the BS array response vector \cite{Balanis2005}. The angles $\psi_{r,b}$ and $\theta_{r,b}$ describe the angle-of-arrival at the RIS and the angle-of-departure at the BS, respectively. We assume that additional multipath components are negligible for this link.

For the channel between RIS $r$ and user $u$ we consider both, a LOS component and additional multipath components modeled as Rayleigh fading. The overall RIS-user channel, denoted by $\boldsymbol{h}_{u,r} \in \mathbb{C}^{M_{\mathrm{RIS}} \times 1}$, therefore follows a Rician fading model
\begin{equation}
\boldsymbol{h}_{u,r} = \gamma_{u,r} \left( \sqrt{\frac{K_{u,r}}{1+K_{u,r}}} \boldsymbol{a}(\theta_{u,r}) + \sqrt{\frac{1}{1+K_{u,r}}} \boldsymbol{g}_{u,r} \right),
\end{equation}
where $\gamma_{u,r}$ is the path gain, $K_{u,r}$ denotes the Rician $K$-factor, $\boldsymbol{a}(\theta_{u,r})$ is the RIS array response vector, and $\boldsymbol{g}_{u,r} \sim \mathcal{CN}(\boldsymbol{0}, \boldsymbol{I})$ models the NLOS component. The angle-of-departure at the RIS is $\theta_{u,r}$. The previously defined path gains $\gamma_{u,b}, \gamma_{r,b}, \gamma_{u,r}$ depend on both distance and line-of-sight conditions.

Each RIS applies a diagonal phase-shift matrix
\begin{equation}
\boldsymbol{\Phi}_r = \mathrm{diag}\!\left(e^{j\phi_{r,1}}, \ldots, e^{j\phi_{r,M_{\mathrm{RIS}}}}\right),
\end{equation}
where the phase shifts are configured to coherently align the LOS components of the RIS-assisted channel. We assume that the random scattering components $\boldsymbol{g}_{u,r}$ vary rapidly over time, which prevents their reliable estimation. Consequently, these components cannot be coherently phase-aligned at the RIS and therefore contribute only incoherently to the received signal. If a RIS is not assigned to the serving BS, its phase shifts are assumed to be random, i.e., $\phi_{r,i} \sim \mathcal{U}(0,2\pi)$.

The aggregate RIS-assisted channel between BS $b$ and user $u$, $\boldsymbol{h}^\mathrm{indirect}_{u,b} \in \mathbb{C}^{M_\text{BS} \times 1}$ is given by
\begin{equation}
\boldsymbol{h}^{\mathrm{indirect}}_{u,b} = \sum_{r=1}^{N_{\mathrm{RIS}}} \left( \boldsymbol{h}_{u,r}^T \boldsymbol{\Phi}_r \boldsymbol{H}_{r,b} \right)^T.
\end{equation}
The total channel $\boldsymbol{h}_{u,b} \in \mathbb{C}^{M_\text{BS} \times 1}$ is then
\begin{equation}
\boldsymbol{h}_{u,b} = \boldsymbol{h}^{\mathrm{direct}}_{u,b} + \boldsymbol{h}^{\mathrm{indirect}}_{u,b}.
\end{equation}

\subsection{Beamforming Model}

We assume strong shadowing of the direct link, such that users are effectively served only via RISs. Consequently, base station beamforming is directed toward RISs and is based solely on the directional channel components represented by the array response vectors of the dominant paths. Rapidly varying NLOS components are not exploited, as they cannot be reliably estimated.

We point beams towards the RISs and we assign power across these beams in a user-specific way. Let $\boldsymbol{f}_{u,d} \in \mathbb{C}^{M_{\mathrm{BS}} \times 1}$ denote the beamforming vector used by the serving base station $d$ for user $u$, including the power allocation, such that $\mathbb{E}[\|\boldsymbol{f}_{u,d}\|^2] = P_d$. 

When a set $\mathcal{R}^{(d)}$ of RISs is assigned to base station $d$, the beamforming vector is constructed as
\begin{equation}
\boldsymbol{f}_{u,d} = \sqrt{\frac{1}{M_{\mathrm{BS}}}} \sum_{r \in \mathcal{R}^{(d)}} \sqrt{P_{u,r,d}} \, \boldsymbol{a}^*(\theta_{r,d}),
\end{equation}
where $\boldsymbol{a}(\theta_{r,d})$ is the array response vector at the base station corresponding to the direction of RIS $r$, $P_{u,r,d}$ denotes the power allocated to user $u$ via RIS $r$, and $\sum_{r \in \mathcal{R}^{(d)}} P_{u,r,d} = P_d$. The power allocation will be defined in Section III. 

If no RIS is assigned to base station $d$, directional information is unavailable. In this case, the base station applies random Gaussian beamforming for the users it serves.

\subsection{Signal Model}

We consider a downlink transmission model in which each BS serves multiple users using orthogonal time-frequency resources. As a result, intra-cell interference is not present. However, inter-cell interference remains present due to simultaneous transmissions from neighboring BSs.

The received signal at user $u$ served by BS $d$ is obtained as
\begin{equation}
y_u = \boldsymbol{h}_{u,d}^T \boldsymbol{f}_{u,d} x_u + \sum_{b \neq d} \boldsymbol{h}_{j_b,b}^T \boldsymbol{f}_{j_b,b} x_j + n_u,
\end{equation}
where $j_b$ denotes the user which is served by BS $b$ at the same time as user $u$ is served by BS $d$, and $n_u \sim \mathcal{CN}(0,\sigma_n^2)$ is additive white Gaussian noise. The transmit symbol intended for user $u$ is denoted by $x_u$ and satisfies $\mathbb{E}[|x_u|^2] = 1$.

\subsection{SINR Model}

The \gls{sinr} for user $u$ served by BS $d$ is expressed as
\begin{equation} \label{eq:instant_sinr}
\mathrm{SINR}_u^{(d)} =
\frac{ \left| \boldsymbol{h}_{u,d}^T \boldsymbol{f}_{u,d} \right|^2 }{ \sigma_n^2
+
\sum_{b \neq d} \frac{1}{|\mathcal{U}^{(b)}|} \sum_{j \in \mathcal{U}^{(b)}} \left| \boldsymbol{h}_{u,b}^T \boldsymbol{f}_{j,b} \right|^2 }.
\end{equation}

The users assigned to base station $b$ are denoted by $\mathcal{U}^{(b)}$. For tractability, we replace the instantaneous inter-cell interference by its average over users in neighboring cells, yielding a scheduling-agnostic interference model. This approximation reflects the fact that channel coding spans many resource elements with potentially varying interferers, and the resulting SINR is interpreted as an effective long-term metric.

The achievable downlink rate of user $u$ served by BS $d$ is then given by
\begin{equation}
r_u^{(d)} = \log_2\!\left(1 + \mathrm{SINR}_u^{(d)}\right).
\end{equation}

\section{SINR and Utility Estimation}

In order to evaluate RIS allocations and guide the auction-based resource assignment, each base station requires an estimate of the achievable performance under a given RIS configuration. Since instantaneous channel state information is not available prior to RIS allocation and configuration, we rely on macroscopic channel parameters and asymptotic properties of large antenna arrays to estimate the SINR and the resulting utility.

\subsection{Macroscopic SINR Estimation}

We approximate the instantaneous received power terms by their respective expected values. For sufficiently large antenna arrays and RISs, this is justified by the law of large numbers.

The estimated SINR of user $u$ served by base station $d$ is expressed as
\begin{equation}
\hat{\mathrm{SINR}}_u^{(d)} =
\frac{ p_{d,u} + p_{c,u} + p_{i,u} }{ \sigma_n^2 + i_{d,u} + i_{i,u} },
\end{equation}
where $p_{d,u}$ denotes the direct signal power, $p_{c,u}$ and $p_{i,u}$ represent the coherent and incoherent RIS-assisted signal components, respectively, and $i_{d,u}$ and $i_{i,u}$ denote direct and RIS-assisted interference. The noise is denoted by $\sigma_n^2$.

\subsection{Signal Power Estimation}

It is justified to split the signal power from \eqref{eq:instant_sinr} into two parts, direct and indirect powers, since we assume a Rayleigh fading link for the direct path. As the Rayleigh fading components are assumed to be unknown for beamforming and RIS configuration, this implies that the indirect RIS-assisted channels are not coherently combined with the direct NLOS channel. 

We will first estimate the direct part. The beamforming vector is statistically independent from the direct channel, because it is matched to the RIS channel. We have:
\begin{equation*}
    \mathbb{E} \left[ \left\| (\boldsymbol{h}_{u,d}^{\text{direct}})^T \boldsymbol{f}_{u,d} \right\|^2 \right] = \mathbb{E} \left[ \left\| \gamma_{u,d} \boldsymbol{g}_{u,d}^T \boldsymbol{f}_{u,d} \right\|^2 \right] =
\end{equation*}
\begin{equation*}
     \gamma_{u,d}^2 \boldsymbol{f}_{u,d}^H \mathbb{E} \bigl[ \boldsymbol{g}_{u,d}^* \boldsymbol{g}_{u,d}^T \bigl] \boldsymbol{f}_{u,d} = \gamma_{u,d}^2 \left\| \boldsymbol{f}_{u,d} \right\|^2 = \gamma_{u,d}^2 P_d = p_{d,u},
\end{equation*}
where $\mathbb{E} \bigl[ \boldsymbol{g}_{u,d}^* \boldsymbol{g}_{u,d}^T \bigl] = \boldsymbol{I}$.

The RIS-assisted signal (the indirect part of the signal from \eqref{eq:instant_sinr}) is the following:
\begin{equation*}
    \mathbb{E} \left[ \left\| (\boldsymbol{h}_{u,d}^{\text{indirect}})^{T} \boldsymbol{f}_{u,d} \right\|^2 \right] = \mathbb{E} \left[ \left\| \boldsymbol{h}_{u,r}^T \boldsymbol{\Phi}_r \boldsymbol{H}_{r,d} \boldsymbol{f}_{u,d}  \right\|^2 \right].
\end{equation*}
We additionally assume asymptotic orthogonality:
\begin{equation*}
    \boldsymbol{a}(\theta_{r,d})^T \boldsymbol{a}(\theta_{s,d})^* \approx 0, \forall s \neq r,\quad  \boldsymbol{a}(\theta_{r,d})^T \boldsymbol{a}(\theta_{r,d})^* = M_\text{BS}.
\end{equation*}
The RIS-assisted signal contains both coherent and incoherent components. The coherent component arises from the line-of-sight parts of the RIS-UE channels that can be phase-aligned by the RISs, while the incoherent component is caused by the non-line-of-sight Rayleigh fading contributions, which cannot be coherently combined as they are assumed to be unknown. After substituting the channel and beamforming expressions, the coherent signal over RIS $r \in \mathcal{R}^{(d)}$ is:
\begin{equation*}
    s_{u,r,d} =  \underbrace{\gamma_{u,r} \gamma_{r,d} k_{u,r} M_\text{RIS} \sqrt{M_\text{BS}}}_{c_{u,r,d}} \underbrace{\sqrt{P_{u,r,d}}}_{m_{u,r,d}},
\end{equation*}
where $k_{u,r} = \sqrt{K_{u,r}/(1+K_{u,r})}$ and $m_{u,r,d}$ is obtained from the power allocation. The total received signal over all allocated RISs is given by:
\begin{equation} \label{eq:indirect_power}
    p_{c,u} = \left( \sum_{r \in \mathcal{R}^{(d)}} c_{u,r,d} m_{u,r,d} \right)^2 = \boldsymbol{m}_{u,d}^T (\boldsymbol{c}_{u,d} \boldsymbol{c}_{u,d}^T) \boldsymbol{m}_{u,d},
\end{equation}
where for $r_i \in \mathcal{R}^{(d)}$: $\boldsymbol{c}_{u,d} = [c_{u,r_1,d}, \ldots, c_{u,r_{\mathcal{R}^{(d)}},d}]$ and $\boldsymbol{m}_{u,d} = [m_{u,r_1,d}, \ldots, m_{u,r_{\mathcal{R}^{(d)}},d}]$.

We maximize \eqref{eq:indirect_power} with respect to the power allocation with the constraint $||\boldsymbol{m}_{u,d}||^2 = P_{d}$. The optimal power allocation is calculated as:
\begin{equation}
    \boldsymbol{m}_{u,d} = \frac{\boldsymbol{c}_{u,d}}{||\boldsymbol{c}_{u,d}||} \sqrt{P_d},
\end{equation}

Due to the asymptotic orthogonality assumption, only the allocated RISs contribute to the effective channel.

Furthermore, the coherent RIS-assisted signal power is given by
\begin{equation}
p_{c,u} = \left( \sum_{r \in \mathcal{R}^{(d)}} s_{u,r,d}\right)^2.
\end{equation}

Regarding the non-coherent power we use similar steps, which results in the incoherent RIS-assisted signal power, given by:
\begin{equation}
p_{i,u} = \sum_{r \in \mathcal{R}^{(d)}} \gamma_{u,r}^2 \gamma_{r,d}^2 \bar{k}_{u,r}^2 M_{\mathrm{BS}} M_{\mathrm{RIS}} P_{u,r,d},
\end{equation}
where $\bar{k}_{u,r} = \sqrt{1/(1+K_{u,r})}$.

\subsection{Interference Power Estimation}

The interference power (analogously to the signal power) consists of two components: direct interference from other base stations and RIS-assisted interference caused by RISs not assigned to the serving base station. From the point-of-view of base station $d$, we consider the beamformers applied at the other interfering base stations as isotropically distributed. This makes sense, as the beamformers applied by the interfering base stations (which point to their assigned RISs) are not correlated with the interference channels.

The direct interference power is given by
\begin{equation}
i_{d,u} = \mathbb{E} \left[ \left\| \boldsymbol{h}_{u,b}^T \boldsymbol{f}_{j_b,b} \right\| ^2 \right] = \sum_{b \neq d} \gamma_{u,b}^2 P_{b}.
\end{equation}
Under these assumptions, the RIS-assisted interference power is expressed as
\begin{equation}
i_{i,u} = \sum_{b \neq d} \sum_{r \notin \mathcal{R}^{(d)}} \gamma_{u,r}^2 \gamma_{b,r}^2 P_{b} M_{\text{RIS}}.
\end{equation}
Because of the isotropic beamforming assumption we do not apply averaging over the users as in \eqref{eq:instant_sinr}. 

The estimated achievable rate of user $u$ served by base station $d$ then follows as
\begin{equation}
\hat{r}_u^{(d)} = \log_2\!\left(1 + \hat{\mathrm{SINR}}_u^{(d)}\right).
\end{equation}

\subsection{Utility Function}

The utility function used for RIS allocation is based on the mean achievable rate of the users served by each base station. For base station $b$, the utility associated with a given RIS allocation $\mathcal{R}^{(b)}$ is defined as
\begin{equation}
\mathrm{Util}^{(b)}(\mathcal{R}^{(b)}) = \frac{1}{|\mathcal{U}^{(b)}|} \sum_{u \in \mathcal{U}^{(b)}} \hat{r}_u^{(b)}.
\end{equation}
This utility captures the average service quality experienced by the users associated with base station $b$.

\subsection{Auction Format}

RIS allocation among base stations is performed using a simultaneously ascending auction, which provides a low-complexity alternative to combinatorial mechanisms such as Vickrey–Clarke–Groves \cite{Nisan2001} while capturing competitive interactions. A similar RIS auction framework was considered in \cite{Schwarz2024}.

The auction proceeds in discrete rounds $t$. In each round, the auctioneer announces a uniform price $p_t$, increased by a fixed increment $\Delta_p$ from the previous round. Each base station $b$ submits a binary bid vector $\boldsymbol{b}_t^{(b)} \in \{0,1\}^{N_{\mathrm{RIS}}}$, where $\boldsymbol{b}_t^{(b)}[r]=1$ indicates willingness to bid for RIS $r$ at price $p_t$.

RISs receiving a single bid are allocated at the current price; RISs with multiple bids remain contested and advance to the next round, while RISs receiving no bids remain unassigned and apply random phase shifts. An activity rule prevents strategic re-entry, i.e., a base station cannot bid for a RIS in round $t$ if it did not bid for it in round $t-1$ \cite{Roughgarden2016}. The activity rule supports the identification of preferences among agents. The auction terminates once no RIS receives multiple bids.

\section{Bidding Strategies}

Let $\mathcal{R}^{(b)}_{t-1}$ denote the set of RISs already allocated to base station $b$ in previous rounds. The set of RISs that remain available at round $t$ is given by
\begin{equation}
\mathcal{R}_t = \{1,\dots,N_{\mathrm{RIS}}\} \setminus \bigcup_b \mathcal{R}^{(b)}_{t-1}.
\end{equation}

Ideally, a base station would evaluate the utility of all possible subsets of remaining RISs; however, this combinatorial evaluation becomes infeasible as the number of RISs grows. We therefore adopt a simplified marginal approach, in which each base station estimates the utility gain of acquiring a single additional RIS, assuming no other RIS is acquired in the same round.

\subsection{Marginal Utility Value Estimation}

The estimated marginal utility value of RIS $r \in \mathcal{R}_t$ for base station $b$ at auction round $t$ is defined as
\begin{equation}
V^{(b)}_t(r) =
\mathrm{Util}^{(b)}\!\left(\mathcal{R}^{(b)}_{t-1} \cup \{r\}\right) -
\mathrm{Util}^{(b)}\!\left(\mathcal{R}^{(b)}_{t-1}\right).
\end{equation}
This value represents the expected improvement in the mean achievable rate of the users served by base station $b$ if RIS $r$ were to be allocated to it.

\subsection{Normalization and Standardization}

While marginal utility values capture the relative desirability of RISs, their absolute scale depends on the channel realization, user distribution, and current allocation. Consequently, the magnitude of $V^{(b)}_t(r)$ can vary significantly across RISs, auction rounds, and training episodes. To obtain a bounded and numerically stable representation suitable for learning-based bidding, we apply a two-step standardization procedure.

First, negative marginal values are clipped using a rectified linear unit (ReLU):
\begin{equation}
\tilde{V}^{(b)}_t(r) = \max\!\left( V^{(b)}_t(r),\, 0 \right),
\end{equation}
such that only RISs expected to yield a performance improvement are considered. A value of zero therefore indicates that no utility gain is anticipated from acquiring RIS $r$ in the current round.

Second, the remaining values are normalized by the maximum positive marginal gain among all available RISs,
\begin{equation}
V^{(b)}_t(r) \leftarrow
\begin{cases}
\dfrac{\tilde{V}^{(b)}_t(r)}
{\max\limits_{r' \in \mathcal{R}_t} \tilde{V}^{(b)}_t(r')},
& \text{if } \max\limits_{r' \in \mathcal{R}_t} \tilde{V}^{(b)}_t(r') \neq 0, \\[10pt]
0, & \text{otherwise}.
\end{cases}
\end{equation}

This normalization maps marginal utility values to the interval $[0,1]$, where one corresponds to the RIS with the highest expected utility gain in the current auction round. The relative ranking of RISs is preserved, while numerical consistency across environments and training episodes is ensured.

\subsection{RL-Based Bidding}

To enable adaptive and fairness-aware bidding, we model the auction as a multi-agent reinforcement learning problem in which each base station acts as an autonomous agent. Through repeated interaction with the auction environment, agents learn bidding strategies that account for both their own utility gains and the relative performance of other base stations.

Unlike purely local strategies, this formulation enables implicit coordination via shared information provided by the auctioneer. In particular, agents are informed of their relative service quality through a fairness-aware weighting mechanism that biases bidding toward weaker-performing cells.

\subsubsection{States}

The complete environment state at auction round $t$ is defined as
\begin{equation}
\mathcal{S}_t = \Bigl( p_t,\; \bigl\{ V^{(b)}_t(r),\; B^{(b)}_t,\; w^{(b)}_t \bigr\}_{\forall b,\, r} \Bigr),
\end{equation}
where $p_t$ denotes the current auction price, $B^{(b)}_t$ is the remaining budget of base station $b$, $V^{(b)}_t(r)$ are the normalized marginal utility values, and $w^{(b)}_t$ is a fairness weight associated with base station $b$.

The fairness weights are computed centrally based on the current utility values of all base stations and are defined as
\begin{equation}
w^{(b)}_t =
\frac{\left(\mathrm{Util}^{(b)} (\mathcal{R}^{(b)}_{t-1})\right)^{\gamma}}
{\sum\limits_{b'} \left(\mathrm{Util}^{(b')} (\mathcal{R}^{(b')}_{t-1})\right)^{\gamma}}
\cdot N_{\mathrm{BS}},
\end{equation}
where $\gamma \geq 0$ controls the strength of the fairness mechanism. For $\gamma = 0$, all base stations are assigned identical weights, whereas larger values of $\gamma$ increasingly emphasize performance differences between base stations. The normalization ensures a unit average fairness weight across base stations, aiming to stabilize the total price expenditure without strictly enforcing constancy.

\subsubsection{Observations}

Each agent operates on an individual observation derived from the global state. The observation available to base station $b$ at round $t$ is given by
\begin{equation}
\mathcal{O}^{(b)}_t = \Bigl( p_t,\; B^{(b)}_t,\; w^{(b)}_t,\; \bigl\{ V^{(b)}_t(r)\; \bigr\}_{\forall r} \Bigr).
\end{equation}

The observation includes the fairness weights, enabling agents to condition their bidding behavior on the relative performance of other cells. This information exchange is mediated by the auctioneer and does not require direct communication between base stations.

To ensure a fixed-length observation vector, the marginal values $V^{(b)}_t(r)$ are set to $-1$ for RISs that are no longer available or for which base station $b$ is inactive due to the enforced activity rule.

\subsubsection{Actions}

At each auction round, each agent selects a binary bid vector
\begin{equation}
\boldsymbol{b}^{(b)}_t \in \{0,1\}^{N_{\mathrm{RIS}}},
\end{equation}
where $\boldsymbol{b}^{(b)}_t[r] = 1$ indicates that base station $b$ places a bid for RIS $r$ at the current price $p_t$, and $\boldsymbol{b}^{(b)}_t[r] = 0$ otherwise.

\subsubsection{Reward Function}

The reward function is designed to encourage agents to bid for RISs that provide high expected utility gains, while discouraging excessive or budget-violating bidding behavior. For base station $b$ at auction round $t$, the reward is defined as
\begin{equation}
r^{(b)}_t = R^{(b)}_{1,t} - \beta w^{(b)}_t \bigl( R^{(b)}_{2,t} + R^{(b)}_{3,t} \bigr),
\end{equation}
where $\beta$ is a constant that controls the overall aggressiveness of bidding \cite{Zan2026}. Rewards are evaluated before the auction outcome, i.e., after each bidding decision instead of only upon winning a RIS, enabling dense and immediate feedback during training. The three reward components are specified as follows.

The first component captures the total expected value of the bids placed by the agent in the current round:
\begin{equation}
R^{(b)}_{1,t} = \sum_{r=1}^{N_{\mathrm{RIS}}} V^{(b)}_t(r)\, \boldsymbol{b}^{(b)}_t[r].
\end{equation}
This term rewards the agent for selecting RISs that are expected to improve its utility.

The second component penalizes the monetary cost of the bids placed in the current round:
\begin{equation}
R^{(b)}_{2,t} = p_t \sum_{r=1}^{N_{\mathrm{RIS}}} \boldsymbol{b}^{(b)}_t[r].
\end{equation}
This term discourages agents from placing unnecessary bids and promotes selective bidding behavior.

The third component introduces an additional penalty if the total bid cost exceeds the remaining budget of the base station:
\begin{equation}
R^{(b)}_{3,t} = 2 \cdot \max\!\left( p_t \sum_{r=1}^{N_{\mathrm{RIS}}} \boldsymbol{b}^{(b)}_t[r] - B^{(b)}_t,\; 0 \right).
\end{equation}
This term explicitly discourages budget violations by penalizing bids that exceed the available budget. The scaling factor emphasizes budget violations relative to regular bidding costs.

Scaling the cost-related terms $R^{(b)}_{2,t}$ and $R^{(b)}_{3,t}$ by the fairness weight $w^{(b)}_t$ biases bidding toward weaker-performing base stations, penalizing aggressive bids from stronger agents while allowing weaker agents to bid more aggressively. This promotes a more balanced RIS allocation while preserving competition.

The proposed fairness mechanism assumes truthful reporting of performance-related information to the auctioneer. Strategic misreporting could introduce vulnerabilities, as a base station claiming lower performance would reduce its own fairness weight and make competing agents more conservative, potentially gaining an advantage. Addressing such behavior would require additional mechanism-design measures, such as verification or incentive-compatible reporting, which are beyond the scope of this work.

\subsection{Implementation}

The environment is implemented using the Gymnasium interface \cite{gymnasium} with multi-agent support provided by PettingZoo \cite{pettingzoo} and SuperSuit \cite{supersuit}. Each training episode corresponds to a complete auction process, starting from the initial price and ending when RISs receive no further bids.

Agents are trained using a policy-gradient-based method, specifically the Proximal Policy Optimization (PPO) algorithm \cite{ppo} as implemented in Stable-Baselines3 \cite{sb3}. An undiscounted return is employed, reflecting the finite-horizon nature of the auction process, where all decisions within an episode contribute equally to the final allocation outcome. Aside from the undiscounted return, all other hyperparameters were kept at the default values of the PPO implementation. Due to the nature of the PPO implementation CPU-based training was utilized to ensure optimal execution speed.

Training is performed episodically across a large number of independent network realizations, where user locations, RIS positions, and channel parameters are randomized between episodes. The implementation supports a variable number of base stations, users, and RISs.

The implementation will be made available upon acceptance of the work at: https://github.com/MartinMarkZan.

\section{Simulations}

In this section, we evaluate the performance of the proposed fairness-aware RIS allocation framework through numerical simulations. The focus is on quantifying the trade-off between system efficiency and user fairness, as well as on illustrating how the proposed mechanism redistributes RIS resources across base stations.

\subsection{Simulation Setup}

We consider a two-base-station scenario with a fixed number of users and RIS elements. The base stations are located at the opposite edges of the region of interest, while the RISs are deployed along the cell edge on a straight line.  The placement of the RISs increases the competition between the two base stations. User locations are generated uniformly within the cell area. One of the base stations is overloaded with users (denoted as BS0 in the figures); on average, it serves approximately three times as many users as the other base station (BS1). While this initial study focuses on a two-base-station scenario with RISs positioned at the cell edge to clearly illustrate fundamental fairness-performance trade-offs, future work will involve more complex network topologies with a larger number of base stations, users, and diverse RIS placements.

\begin{figure}[t]
\centerline{\includegraphics[width=\columnwidth]{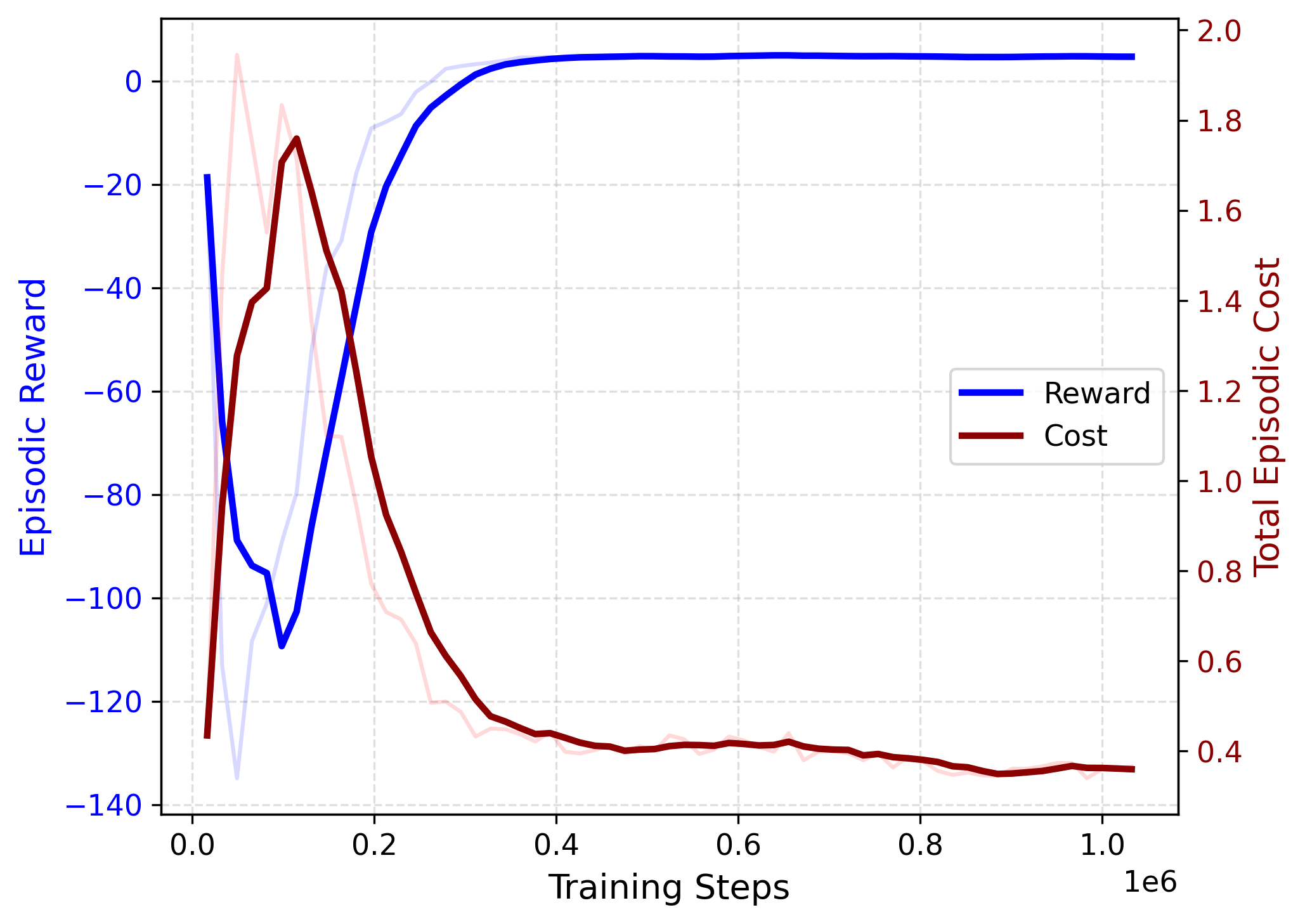}}
\caption{Convergence of the episodic reward (left axis) and total auction cost (right axis) during training. Solid lines represent moving-average smoothed curves (window size=5), while semi-transparent lines show the raw data.}
\label{fig:convergence}
\end{figure}

As illustrated in Fig.~\ref{fig:convergence}, the reinforcement learning agents converged to stable reward values during training, indicating reproducible learning behavior. The curves show that agents reliably discover effective bidding policies that maximize gains while maintaining stable budget utilization, confirming robustness. During evaluation, we used 200 macroscopic realizations (user positions and large-scale path gains) and 20 independent microscopic fading realizations per macroscopic setup to aggregate the results.

\begin{table}[t]
\caption{Simulation Parameters}
\begin{center}
\begin{tabular}{|c|c|}
\hline
Carrier frequency & $f_c = 26\,\text{GHz}$ \\
\hline
Number of base stations & $N_\text{BS} = 2$ \\
\hline
Number of base station antennas & $M_\text{BS} = 50$ \\
\hline
Number of users & $N_\text{UE} = 20$ \\
\hline
Number of RISs & $N_\text{RIS} = 10$ \\
\hline
Number of RIS elements & $M_\text{RIS} = 250$ \\
\hline
Transmit power per subcarrier & $P = 100\,\text{mW}$ \\
\hline
Subcarrier bandwidth & $15\,\text{kHz}$ \\
\hline
AWGN noise power spectral density & $-174\,\text{dBm/Hz}$ \\
\hline
Noise figure & $6\,\text{dB}$ \\
\hline
Path-loss exponent under LOS (NLOS) & $2 \ (4.5)$ \\
\hline
$K$-factor under LOS (NLOS) & $100 \ (3)$ \\
\hline
Distance-dependent LOS-probability & $p_\text{LOS}(d) = e^{-d/25}$ \\
\hline
Shadow fading variance & $10\,\text{dB}$ \\
\hline
Auction initial price & $p_0 = 0.05$ \\
\hline
Auction price increment & $\Delta p = 0.05$ \\
\hline
Budget & $B^{(b)}_0 = 1$ \\
\hline
\end{tabular}
\label{tab:sim_params}
\end{center}
\end{table}

\begin{figure}[t]
\centerline{\includegraphics[width=\columnwidth]{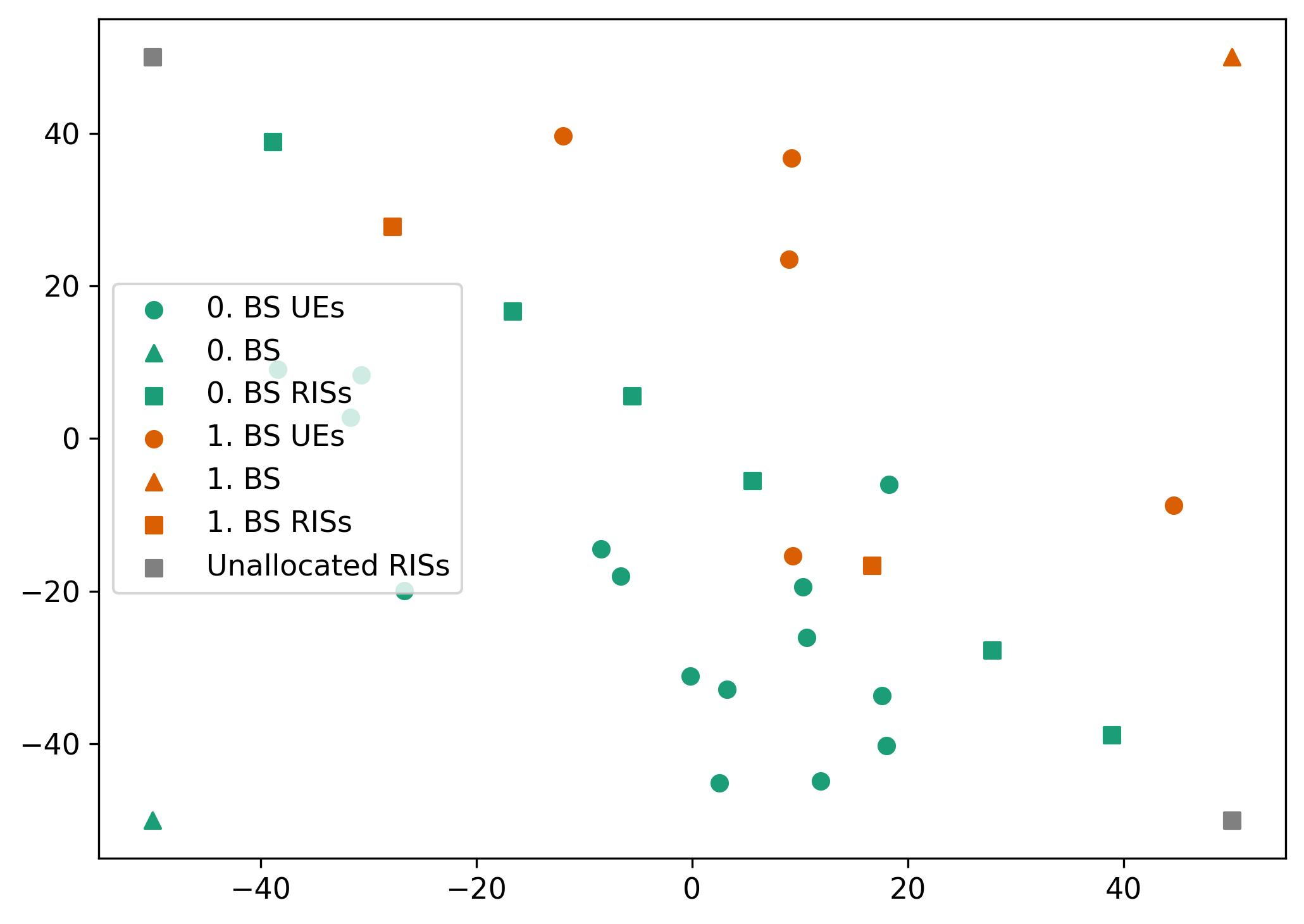}}
\caption{Representative network realization for $\gamma = 0.2$, showing the
locations of the two base stations, users, allocated RISs, and unassigned RISs.}
\label{fig:geometry}
\end{figure}

Table~\ref{tab:sim_params} summarizes the main simulation parameters. The geometry is shown in Fig.~\ref{fig:geometry} for a random snapshot of user positions.

\subsection{Efficiency-Fairness Trade-off}

\begin{figure}[t]
\centerline{\includegraphics[width=\columnwidth]{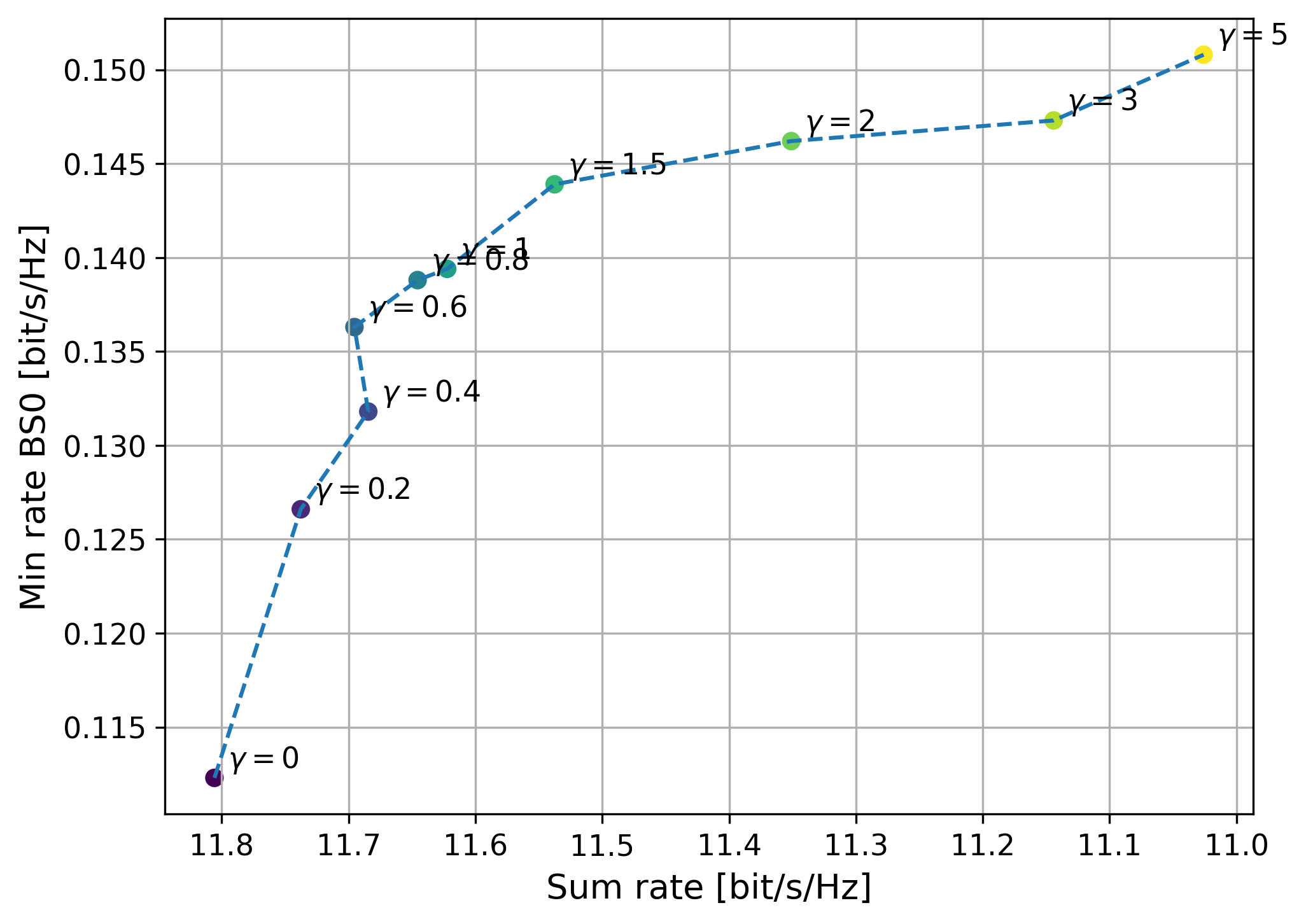}}
\caption{Trade-off between sum rate and the minimum user rate of the overloaded
base station (BS0). Each point corresponds to a model with a different
value of the fairness strength $\gamma$. } 
\label{fig:min_vs_sum}
\end{figure}

Fig.~\ref{fig:min_vs_sum} visualizes the trade-off between system efficiency and fairness using the sum rate versus the minimum user rate of the overloaded base station.

As $\gamma$ increases, the operating point moves along a Pareto-like frontier: the minimum rate of BS0 improves by approximately $34\%$, while the sum rate of the two base stations decreases only moderately (less than $7\%$ over the considered range). This confirms that the proposed mechanism is able to substantially improve the performance of the worst-served users without causing a severe loss in overall system throughput.

\subsection{Fairness Evaluation via Atkinson Index}

\begin{figure}[t]
\centerline{\includegraphics[width=\columnwidth]{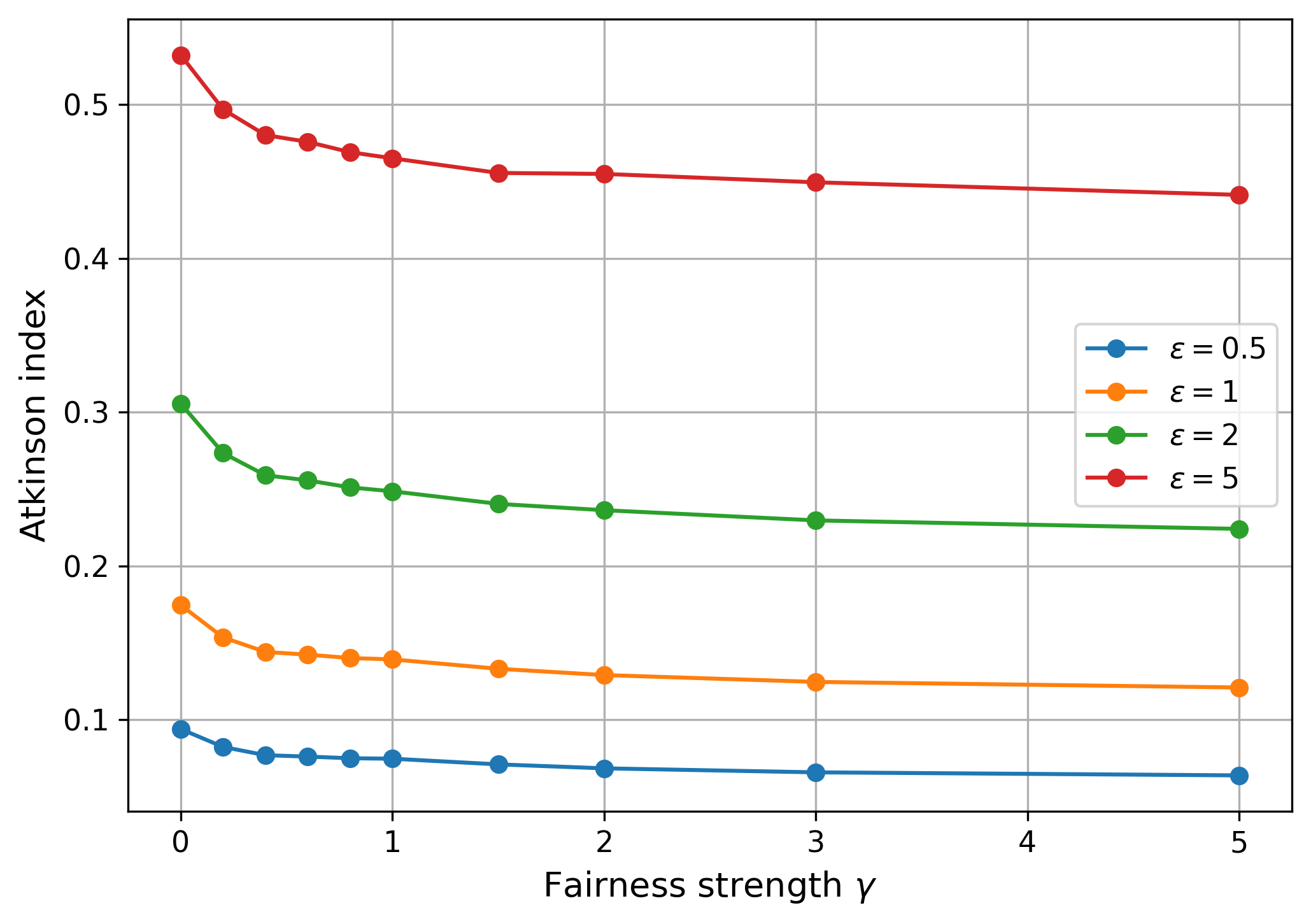}}
\caption{Atkinson inequality index as a function of the fairness strength $\gamma$
for different values of the sensitivity parameter $\epsilon$, which controls the
emphasis on low-rate users.}
\label{fig:atkinson}
\end{figure}

To quantify fairness more systematically, Fig.~\ref{fig:atkinson} reports the Atkinson inequality index as a function of the fairness strength $\gamma$. The index is defined as
\begin{equation}
A_\varepsilon (y_1,\ldots,y_N) = 1 - \frac{E_\varepsilon}{\mu},
\end{equation}
where
\begin{equation*}
E_\varepsilon =
\begin{cases}
\left(
\dfrac{1}{N} \sum_{i=1}^{N} y_i^{\,1-\varepsilon}
\right)^{\!\frac{1}{1-\varepsilon}},
& 0 \leq \varepsilon \neq 1, \\[6pt]
\left(
\prod_{i=1}^{N} y_i
\right)^{\!\frac{1}{N}},
& \varepsilon = 1, \\[6pt]
\min(y_1,\ldots,y_N),
& \varepsilon = +\infty,
\end{cases}
\end{equation*}
and $\mu$ is the mean of the input values. The Atkinson index takes values in $[0,1]$, where smaller values indicate more equal rate distributions. 

For all considered $\epsilon$, increasing $\gamma$ consistently reduces the inequality index, confirming that the proposed framework improves fairness across users. Larger values of $\epsilon$ result in higher Atkinson indices, since the metric places greater emphasis on the worst-served users and penalizes residual disparities more strongly. The monotonic decrease of all curves with $\gamma$ demonstrates that the fairness improvement is robust with respect to the chosen fairness sensitivity.

\subsection{RIS Allocation Behavior}

\begin{figure}[t]
\centerline{\includegraphics[width=\columnwidth]{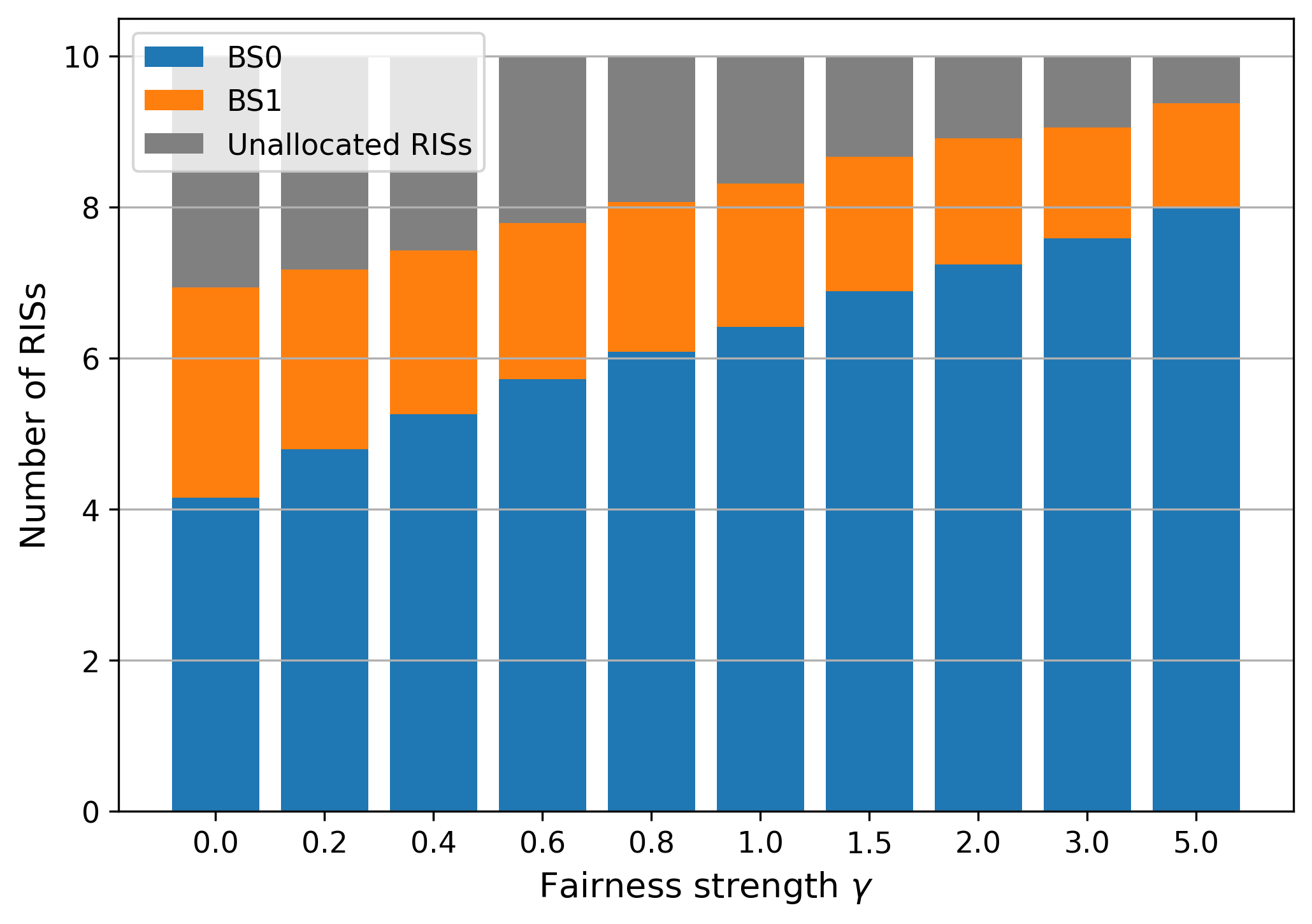}}
\caption{Increasing $\gamma$ shifts RIS resources from BS1 to the overloaded BS0, while also decreasing the number of unallocated RISs due to more aggressive bidding behavior. }
\label{fig:ris_dist}
\end{figure}

Fig.~\ref{fig:ris_dist} shows the average RIS allocation as a function of the fairness parameter $\gamma$. The figure reports the number of RISs assigned to BS0, BS1, and those remaining unallocated.

As $\gamma$ increases, RIS resources are progressively shifted from BS1 to the overloaded BS0, which directly explains the observed improvement in the minimum rate of BS0. At the same time, the number of unallocated RISs decreases, indicating more aggressive bidding behavior by the weaker-performing base station and a less competitive auction.

For the considered operating point, the total price spent by the two base stations remains approximately constant across $\gamma$. However, this behavior is not guaranteed in general. Additional experiments with different cost-scaling parameters ($\beta$) show that stronger fairness pressure can also lead to increased overall expenditure. This highlights an inherent interaction between fairness objectives and economic efficiency, which can be controlled through appropriate reward design.

\section{Conclusion}

In this work, we investigated auction-based allocation of reconfigurable intelligent surfaces in asymmetric multi-cell networks with a focus on fairness-aware resource distribution. RISs were modeled as shared infrastructure and dynamically allocated through a simultaneous ascending auction, enabling scalable allocation in competitive cell-edge scenarios. To address performance imbalances caused by uneven user distributions, we proposed a cooperative multi-agent reinforcement learning framework that integrates a performance-dependent fairness mechanism into the bidding process.

Simulation results show that the proposed approach significantly improves the performance of the worst-served users, while maintaining competitive sum-rate performance. By adjusting the fairness parameter, the trade-off between data rate and equitable resource allocation can be explicitly controlled. These findings demonstrate the effectiveness of combining auction-based mechanisms with cooperative reinforcement learning for fair and efficient RIS utilization in future wireless networks.

While the proposed framework demonstrates strong performance in moderate-sized scenarios, extending the approach to large-scale networks with a higher number of base stations, users, and RISs remains an important research direction. Additionally, other auction formats, such as sealed-bid mechanisms or dynamic pricing schemes, could be investigated to further improve efficiency or fairness under different deployment assumptions. Future work can also consider non-stationary environments with time-varying users.

\end{document}

%% file: acronyms.tex
\newacronym{ris}{RIS}{reconfigurable intelligent surface}
\newacronym{bs}{BS}{base station}
\newacronym{ue}{UE}{user}
\newacronym{los}{LOS}{line-of-sight}
\newacronym{nlos}{NLOS}{non-line-of-sight}
\newacronym{sinr}{SINR}{signal-to-interference-plus-noise ratio}

\newacronym{rl}{RL}{reinforcement learning}

\newacronym{comp}{CoMP}{coordinated multipoint transmission}

\newacronym{drl}{DRL}{deep reinforcement learning}